\documentclass[conference, letterpaper]{IEEEtran}
\usepackage[T1]{fontenc}
\usepackage{amsmath}
\usepackage{amsfonts}
\usepackage{indentfirst}
\usepackage{picture}
\usepackage{amssymb}
\usepackage{url}
\usepackage{graphicx}
\usepackage[table]{xcolor}
\usepackage[numbers]{natbib}
\usepackage{array}
\usepackage{threeparttable}
\usepackage{longtable}
\usepackage{hyperref}
\begin{document}

\title{Service Function Chaining Resource Allocation:\\ A Survey}
\author{\IEEEauthorblockN{Yanghao Xie, Zhixiang Liu, Sheng Wang, Yuxiu Wang}\\
\IEEEauthorblockA{Key Laboratory of Optical Fiber Sensing and Communication, Education Ministry of China\\University of Electronic Science and Technology of China\\Email: yanghao.xie@std.uestc.edu.cn}}

\maketitle
\begin{abstract}
	Service Function Chaining (SFC) is a crucial technology for future Internet. It aims to overcome the limitation of current deployment models which is rigid and static. Application of this technology relies on algorithms that can optimally mapping SFC to substrate network. This category of algorithms is referred as ``Service Function Chaining Resource Allocation (SFC-RA)'' algorithms or ``VNF placement (VNFP)'' algorithms. This paper presents a survey of current researches in SFC-RA algorithms. After presenting the formulation and related problems, several variants of SFC-RA problem are summarized. At last, we discussed several future research directions.
\end{abstract}


\section{Introduction}
\label{Introduction}
\IEEEPARstart{w}{ith} the fast development of Internet and network services, more and more middleboxes are deployed in networks for technical reasons, value-add reasons, etc. A recent paper shows that the number of middleboxes is comparable to the number of routers in an enterprise network \cite{SherryRatnasamyAt2012}. However, middleboxes means high \emph{Capital Expenditures (CAPEX)} and \emph{Operational Expenditures (OPEX)}, moreover, deployment or re-deployment of middleboxes needs expertise which increases OPEX and decreases flexibility. Generally, middleboxes are referred as service functions, network functions, or functions, we use those terminologies interchangeably in this paper.\\
\indent
Other problems arise as the fact that most often a flow is required to pass through a sequence of middleboxes in a particular order which is typically referred to as Service Function Chaining (SFC) \cite{QuinnNadeau2015}. For example, current service function chaining deployment model is topology-dependent and device-specific; therefore, adding, deleting, and modifying service function chains can be cumbersome and error-prone, even worse, those tasks could be incompletable. All those features show the incompetent of current deployment model \cite{QuinnNadeau2015}.\\
\indent 
Due to the emerging technologies of Network Function Virtualisation (NFV) and Software Defined Network (SDN), future Internet will be a virtualized environment in which Virtual Network Functions (VNFs) are deployed on Network Function Virtualization Infrastructure (NFVI) and VNFs can be deployed on-demand and customized by the VNF providers \cite{MijumbiSerratGorrichoEtAl2015}. Therefore, future SFC deployment model will follow the philosophy of NFV and SDN. Especially, Service Function Chaining literally means an ordered list of instances of network functions that traffic traverses through, current implementation of service function chaining is typically based on VLAN, policy based routing, individual VRF (Virtual Routing and Forwarding), etc. In this paper, we use the terminology ``Service Function Chaining'' following the convention of IETF Service Function Chaining Working Group (IETF SFC WG), where SFC means a novel service chain deployment model in NFV and SDN context.\\
\indent 
There is an urgent demand for dynamic, elastic, and flexible service chaining deployment model. This has attracted attention from both industry and academic community. OpenDayLight \cite{2016} has created a project about SFC, and the project aims to provide the infrastructure (chaining logic, API) needed for ODL to provision a service chain in the network and an end-user application for defining such chains. Similarly, OpenStack \cite{2016a} introduces an extension to provide APIs and implementations to support SFC. In addition, IETF SFC WG has several drafts demonstrating SFC use cases in data center \cite{KumarTufailMajeeEtAl2016} and mobile network \cite{NapperHaeffnerStiemerlingEtAl2016}, and RFCs discussing architecture \cite{HalpernPignataro2015} and problem statement \cite{QuinnNadeau2015}. Moreover, \citet{QuinnElzur2016} propose an encapsulation which is named as \emph{Network Service Header} to implement SFC. And there are a lot of literature about SFC, e.g., recovery from failure \cite{LeeShin2015}, resource allocation \cite{LukovszkiSchmid2014}, service function specification \cite{Po-WenYu-ChengChin-Laung2015}.\\
\begin{figure}[t]
	\centering	
	\begin{picture}(250,100)(0,0)
	\put(0,0){\dashbox(250,100)[bl]}
	\put(5,5){SFC-enable Domain}	
	\put(-5,50){\vector(1,0){28}}
	\put(2,55){flows}	
	\put(25,25){\dashbox(55,50)}
	\put(25,35){\shortstack{Service\\Classification\\Function}}	
	\put(85,50){\vector(1,0){55}}
	\put(85,55){\shortstack{SFC\\Encapsulation}}	
	\put(145,25){\dashbox(100,50)}
	\put(150,35){\shortstack{Service\\Function\\Path}}	
	\put(190,40){\dashbox(20,20){SF1}}	
	\put(210,50){$...$}	
	\put(220,40){\dashbox(20,20){SFn}}	
	\put(240,50){\vector(1,0){20}}
	\end{picture}
	\caption{Service Function Chain Architecture}
	\label{Service-Function-Chain-Architecture}
\end{figure}
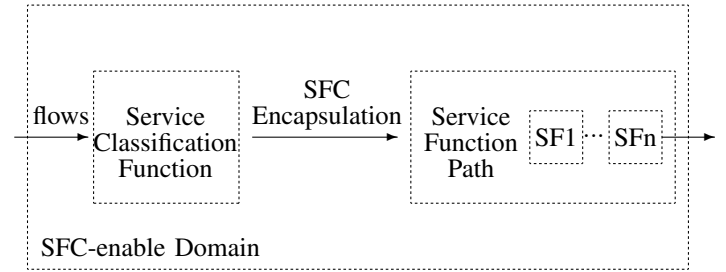
\indent
The SFC Architecture is built out of four logical building blocks: Classifiers, Service Function Forwarders (SFFs), the Service Functions (SFs), and SFC proxies. \figurename{ \ref{Service-Function-Chain-Architecture}} illustrates the basic procedure of flows that traverse the SFC-enable domain (the SFC proxies are not included). When a flow gets into a SFC-enable domain, it will be classified, in order to define which Service Function Path (SFP) the flow should traverse. Then the flow will be encapsulated, which enables SFP selection and the sharing of metadata information if necessary. After the encapsulation, the flow will traverse through all the service functions that are defined by the encapsulation. After all the works are finished, the encapsulation will be removed, and the flow will depart the SFC-enable domain and continue its transmission \cite{HalpernPignataro2015}.\\
\indent
The problem of mapping service functions in NFVI is the main challenge in SFC implementation. By dynamic mapping VNFs onto physical hardware, the advantages obtained from existing hardware service can be maximized. Optimally dynamic resource allocation, which is the main feature of future networks, is crucial to provide on-demand, cost-saving, and environmental-friendly network services.\\
\indent
This calls for efficient algorithms to determine on to which NFVI Point of Present (NFVI-PoP) VNFs are placed, and how VNF migration is implemented due to failures, load balancing, etc. The work of resource allocation in SFC is closely related to Virtual Network Embedding (VNE) and can always be formulated as an optimization problem, related examples are \cite{MehraghdamKellerKarl2014,AddisBelabedBouetEtAl2015,MoensDeTurck2014,GuptaHabibChowdhuryEtAl2015}.\\
\indent For example, \citet{MehraghdamKellerKarl2014} propose a Mixed Integer Quadratically Constrained Program (MIQCP), and \citet{AddisBelabedBouetEtAl2015} present a Mixed Integer Linear Programming (MILP), while \citet{MoensDeTurck2014} formulate the problem as an Integer Linear Program (ILP) . Similarly, \citet{GuptaHabibChowdhuryEtAl2015} give an ILP formulation without an algorithm.\\
\indent However, the SFC-RA problem can be reduced to two well known \emph{NP-hard} optimization problems -- the \emph{facility location problem} and the \emph{generalized assignment problem} (GAP). Therefore, the SFC-RA problem is NP-hard, too \cite{CohenLewin-EytanNaorEtAl2015}, hence intractable for big instances. Therefore, heuristic or approximation algorithms are needed, e.g., \cite{,BariChowdhuryAhmedEtAl2015, CohenLewin-EytanNaorEtAl2015, XiaShirazipourZhangEtAl2014}.\\
\indent For example, \citet{BariChowdhuryAhmedEtAl2015} leverage ILP and propose a heuristic solution based on Viterbi algorithm, which provides solutions that are within 1.3 times of the optimal solution. Furthermore, \citet{CohenLewin-EytanNaorEtAl2015} show that the SFC-RA problem introduces a new type of optimization problems, and provide near optimal approximation algorithm that guarantees a placement with theoretically proven performance. \citet{XiaShirazipourZhangEtAl2014} formulate the problem of VNF placement in packet/optical datacenters as Binary Integer Program (BIP) and propose a heuristic algorithm, however the formulation and algorithm are not fit for general cases. \\
\indent
A recent survey by \citet{XinChen2016} presents network function orchestration frameworks, and particularly the network function placement strategies. In addition, the authors compare different frameworks and present the advantages and disadvantages of different VNF placement approaches. This paper goes beyond what the aforementioned paper provides: The SFC-RA problem is discussed in all its variants and current approaches proposed by academic community are categorized, and a more comprehensive survey of literature is presented. Moreover, we discuss two well-know problems that are closely related to the SFC-RA problem, from which are worth drawing inspiration. In addition, we present a formal and basic formulation of the SFC-RA problem, which captures the generalities of all the variants, and this formulation can be used as a guide to investigate the SFC-RA problem. Future research directions are presented, too.\\
\indent
The remainder of this paper is organized as follows: Section \ref{Basic-mathmatical-model-and-problem-defination} formulates the SFC-RA problem and presents several optimization strategies. Section \ref{Related-problems} presents two similar problems that are closely related to SFC-RA problem. Several variants of SFC-RA are discussed in Section \ref{Variants-of-SFC-RA-problem}. Section \ref{Future-research-directions} discusses future research directions. We conclude this paper in Section \ref{Conclusion}.\\

\section{Basic formulation and optimization strategies}
\label{Basic-mathmatical-model-and-problem-defination}
The application of NFV introduces the problem of how the virtual resource should be realized by NFVI, and in SFC terminology, how the service function chains be placed. An amount of papers have presented scenarios-based formulations for SFC-RA problem. In this section, a basic formulation of SFC-RA is proposed. The formulation is based on the work of \citet{BariChowdhuryAhmedEtAl2015}. Especially, we do not consider mapping \emph{virtual links} between two VNFs, we discuss this in Section \ref{SFC-Resource-Allocation}.
\begin{table}[t]
	\renewcommand{\arraystretch}{1.3}
	\caption{terminology used in formulation}
	\label{terminology-used-for-formulation}
	\centering
	\footnotesize
	\rowcolors*{2}{white}{black!10}
	\resizebox{\columnwidth}{!}{
		\begin{tabular}{l l}
			\hline
			\noalign{\smallskip}
			Term&Description\\
			\hline
			\noalign{\smallskip}
			$G=(S, L)$&Physical network consisting of nodes $S$ and links $L$\\
			$R$&A set represents resources provided by each node\\
			$c_n^r\in \mathbb{R} ^+, \forall r\in R$& Resource capacity of node $s\in S$\\
			$Q$&A set represents different VNF types\\
			$\kappa_q^r\in \mathbb{R^+} (\forall r\in R)$&Resource requirement of VNF type $q\in Q$\\
			$\beta ^q\in \mathbb{R^+}$&Process capacity of VNF type $q$\\
			$T$&A set represents service requests\\
			$G^t=(N^t,L^t)$&Service Request consisting of nodes $N^t$ and links $L^t$\\
			$\beta ^t\in \mathbb{R^+}$&Bandwidth demand of request $t$\\
			$P$&A set represents VNFs needed to be placed\\
			$f: P\rightarrow S$&A function maps VNFs to physical network\\
			$g^t: N^t\rightarrow P, \forall t$&A function maps service requests to VNFs\\
			\noalign{\smallskip}
			\hline 
		\end{tabular}
	}
\end{table}
\subsection{Physical network}
Physical network is where the VNFs instantiate. In virtual environment, software and hardware are decoupled, which means the network functions are independent of the physical elements as current situation, and we can change the software to realize different service functions on one single physical element. Therefore, a physical network element can be considered as a building block that has the capacity of process, storage, communication, etc.\\
\indent 
The physical network is modeled as an undirected graph $G=(S, L)$, where $S$ and $L$ denote the set of nodes and links, respectively. It is assumed that all the nodes are NFVI-PoPs. However, the actual situation may be different as not all the nodes in a network are virtualized, and we do so for simplicity.\\
\indent
Set $R$ denotes resources (CPU, Memory, storage, etc.) that are provided by each node. The resource capacity of node $s\in S$ is denoted as $c_s^r\in \mathbb{R} ^+, \forall r\in R$.
\subsection{Virtualised Network Functions (VNFs)}
A VNF is a virtual version of network function which is commonly known as \emph{middlebox} in current non-virtualised network. For example, technical-reasons network elements, e.g., Network Address Translation (NAT), Fire Wall (FW), Dynamic Host Configuration Protocol (DHCP); value-added network elements, e.g., Malware Detection (MWD), Lawful Interception (LI); and mobile network elements, e.g., TCP Optimization, Video Optimizers, etc \cite{NapperHaeffnerStiemerlingEtAl2016,ISG2014-12a}.\\
\indent
There are different kinds of VNFs in a network, and different kinds of VNFs need different kinds of resources and different quantity of resources. For example, a NAT do not need much computing resource, however, a DPI may need much more computing resource. Set $Q$ denotes different VNF types, and $\kappa_q^r\in \mathbb{R^+}, \forall r\in R$ denotes the resource requirement of VNF type $q\in Q$. The process capacity of VNF type $q$ is denoted as $\beta ^q\in \mathbb{R^+}$ (in Mbps).
\subsection{Service requests}
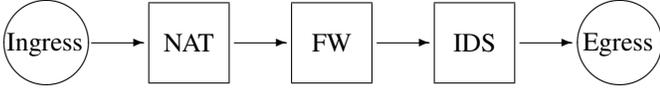
\begin{figure}[t]
	\centering
	\begin{picture}(250,70)(0,10)
	\put(15,35){\circle{30}}
	\put(0,32){Ingress}
	\put(32,35){\vector(1,0){20}}
	\put(54,20){\framebox(30,30){NAT}}
	\put(86,35){\vector(1,0){20}}
	\put(108,20){\framebox(30,30){FW}}
	\put(140,35){\vector(1,0){20}}
	\put(162,20){\framebox(30,30){IDS}}
	\put(194,35){\vector(1,0){20}}
	\put(232,35){\circle{30}}
	\put(218,32){Egress}
	\end{picture}
	\caption{Service Request Model}
	\label{Service-Request-Model}
\end{figure}

On the way to the destination, flows must be steered through an ordered list of instances of network functions. For example, a chain of NAT, FW, and IDS, depicted by \figurename{ \ref{Service-Request-Model}}. The set of enabled service function chains stands for the operators' services and is build according to service agreements between operators and end uses with respect to network policies.\\
\indent
Set $T$ represents service requests. A service request is a path from source to destination that has an ordered list of service functions on it. We model a service request as a directed graph $G^t=(N^t,L^t), \forall t\in T$, where $N^t$ denotes the source, destination, and VNFs, while $L^t$ denotes the \emph{virtual links} between those VNFs. The bandwidth demand of service request $t$ is denoted as $\beta ^t\in \mathbb{R^+}$ (in Mbps).

\subsection{SFC Resource Allocation}
\label{SFC-Resource-Allocation}
In the SFC-RA problem, we are given a physical network, VNF specifications, and a set of service requests, therefore, the SFC-RA has three steps:
\begin{itemize}
	\item[1] Calculate an optimal number of needed VNF types, and all the VNFs that should be instantiated compose a set which is denoted as $P$. In order to simplify the notation, VNF $p\in P$ also denotes the type of $p$;
	\item[2] $f: P\rightarrow S$ s.t. $\forall s\in S, \forall r\in R: \sum_{p|f(p)=s} \kappa_{p}^{r} \leq c_s^r$.\\ Place VNFs to physical nodes such that the demand of VNFs do not exceed the capacity of physical nodes;
	\item[3] $g: N^t\rightarrow P, \forall t$ s.t. $\forall p\in P: \sum_{t|g(N^t)=p}\beta ^t\leq \beta ^p$.\\ Assign service requests to VNFs such that the demand of service requests do not exceed the capacity of VNFs.
\end{itemize} 

We divide the optimization process into three steps, however, the three steps are not independent. Actually, we may not follow the oder presented. For example, different VNF placement and service request assignment can change the number of $P$, and we can obtain all solutions of the three steps simultaneously, according to solving a \emph{Mathematical Program}. We do so to make the SFC-RA problem more vivid and easy to understand.\\
\indent
Some literature assumes that the the path between two VNFs is determined by the physical location of two VNFs, which is reasonable in current network paradigm. In addition, a \emph{virtual link} may consist of several \emph{physical links}, which complicates our basic formulation. Therefore, we do not consider mapping \emph{virtual links} in our formulation for simplicity. However, with the flexibility of Software Defined Network (SDN), some literature manipulate the routing paths between VNFs, which jointly optimize the placement and routing problem in SFC-RA problem \cite{XinChen2016}. Researchers should consider this if necessary.

\subsection{Optimization approaches}
The SFC-RA problem is NP-hard \cite{CohenLewin-EytanNaorEtAl2015}. Therefore, for large problem sizes (i.e. large service chain and physical network size), the cost to solve this problem turn into unaffordable. Considering this hardness, several kinds of appraoches have been used to solve SFC-RA problem. \emph{Exact solutions} find global optimal solutions, however, they always suffer from large problem size. Therefore, exact solutions are commonly used to solve small instances and present an optimal bound reference for heuristic solutions. \emph{Approximation algorithms} find approximate solutions that quality and run-time bounds can be provable. \emph{Heristic algorithms} try to exploit problem-specific knowledge and for which we have no guarantee that they find the optimal solution \cite{Rothlauf2011}. A \emph{metaheuristic} is formally defined as an iterative generation process which guides a subordinate heuristic by combining intelligently different concepts for exploring and exploring the research space, learing strategies are used to structure information in order to find efficiently near-optimal solutions \cite{OsmanLaporte}.
\subsubsection{Exact and approximation solutions}SFC-RA problem can always be formulated as \emph{Optimization Problem}, more exactly, Integer Linear Programming (ILP) or Mixed Integer Linear Programming (MILP), etc. ILP are typically  NP-hard, however, we can use several exact algorithms to solve ILP, e.g., \emph{branch and bound}, \emph{dynamic programming}, \emph{cutting plane methods}, etc. \emph{Approximation solutions} give a trade-off between optimal solution and algorithm complexity, therefore, a polynomial time algorithm can be found according to compromising of optimality.\\
\indent
For example, \citet{TalebBagaaKsentini2015} use ILP to formulate VNF placement in mobile network, and use CPLEX, MATLAB, and CVX to solve the problem. \citet{CohenLewin-EytanNaorEtAl2015} give an ILP formulation, and present an approximation algorithm with proven performance and bound by reducing the problem to \emph{Gneralized Assignment Problem (GAP)}.
\subsubsection{Heuristic solutions}Service chain deployment are supposed to be low delay or real-time, therefore, fast heuristics are preferred in SFC-RA. Therefore, threre are huge quantity of heuristic-based solution for SFC-RA problem.\\
\indent
For example, \citet{BariChowdhuryAhmedEtAl2015} propose a dynamic programming based heuristic to solve large instances, which provides solutions within 1.3 times of the optimal solution obtained by solving ILP using CPLEX. \citet{MohammadkhanGhapaniLiuEtAl2015} give a MILP formulation and a heuristic algorithm that solve the problem incrementally, which can solve the problem for incoming flows without impacting existing flows.
\subsubsection{Metaheuristic solutions}
Metaheuristic algorithms are another way to overcome the hardness of SFC-RA problem. A lot of metaheuristics can be used to find better solutions, e.g., \emph{simulated annealing}, \emph{tabu search},  \emph{genetic algorithms}, etc.\\
\indent
For example, \citet{BouetLeguayConan2013} consider the problem of dynamic deploy Deep Packet Inspection (DPI) in NFV environment. The authors propose a method based on \emph{genetic algorithms}, which provides a tradeoff between the number of engines and the network load to minimize the global cost of the deployment. \citet{MijumbiSerratGorrichoEtAl2015a} consider the placement and scheduling of network functions in NFV. Then a greedy algorithm and a tabu search-based algorithm are proposed to solve the problem effectively.

\section{Related problems}
\label{Related-problems}
Appropriate resource allocation is a very old problem. And it has an amount of instances in different disciplines, e.g., economics \cite{CohenCyertothers1965}, wireless networks \cite{GeorgiadisNeelyTassiulas2006}, etc. In this section, we present two resource allocation problems in computer networks, which are closely related to SFC-RA problem, i.e., VM placement in cloud computing and Virtual Network Embedding. We can draw inspiration from the two well-studied problems. We also discuss the similarities and differences between SFC-RA problem and each of them.

\subsection{Virtual machine placement}
According to the definition of National Institute of Standards and Technology (NIST), Cloud Computing is "a model for enabling ubiquitous, convenient, on-demand network access to a shared pool of configurable resources (e.g., network, servers, storage, applications, and services) that can be rapidly provisioned and released with minimal management effort or service provider interaction \cite{PeterTimothy2011}." In last few decades, Cloud Computing is regarded as one of the most promising technologies in computer science \cite{BuyyaYeoVenugopalEtAl2009}.

\subsubsection{Introduction to virtual machine placement}Users are provided with resources, e.g., computing, storage, network, and so on, from the cloud data center as a service, however, users do not own the resources. Therefore, virtualisation is introduced into cloud computing, in which users' requests are implemented on Virtual Machines (VMs), and multiple VMs are implemented on a Physical Machine (PM). Inside a PM, a layer of software, which is called \emph{hypervisor}, controls all the VMs. Due to the dynamic feature of the Internet, resources required by VMs vary with time. Therefore, VM migration is introduced to overcome this problem in order to guarantee the Service Level Agreement (SLA). For example, when traffic grows, the VM process the traffic may be migrated to another VM that has enough resources. VM migration consists of four steps. First, selecting the PM from which the VM is migrated; second, selecting the VM for the migration; third, selecting the PM that the VM will be placed; four, transfering the VM \cite{GuptaPateriya2014}.\\
\indent
The third step that is selecting a suitable PM that can hosts the VM, which is also called VM placement, is a challenging task, because the performance of the cloud computing is directly impacted by VM placement. Due to the importance of VM placement, a huge amount of VM placement approaches have been proposed in the literature \cite{MengPappasZhang2010,SilvaFonseca2015,ChaseKaewpuangWenEtAl2014,JiangTianSangtaeEtAl2012}. For example, \citet{MengPappasZhang2010} use traffic-aware VM placement to improve the network scalability in data center. The authors formulate the VM placement problem as an optimization problem and prove its hardness, and propose a two-tier approximation algorithm to overcome very large problem sizes efficiently.\\
\indent
Generally, VM placement is the process of selecting the most appropriate physical machines for the virtual machines. According to \cite{masdarinabaviahmadi2016}, common objectives of VM placement are maximizing resource utilization, reliability and availability, etc. In addition. there are several variants of VM placement, e.g., dynamic placement, multi-clouds placement, etc.
\subsubsection{Relationship between SFC-RA and VM placement}
The two problems have the similarity that they both try to place virtual objects on appropriate location, which can be regard as two variants of resource allocation problem. However, different scenarios result in different problems.\\
\indent
In general, SFC originates from telecommunication industry, hence the performance and reliability requirement of network functions are carrier-grade. Therefore, the performance of SFC should be carrier-class, which means that the deployment of SFC should meet or exceed \emph{five nine} high availability standards, and provide very fast fault recovery through redundancy \cite{GreeneLancaster2007}. However, cloud computing has less performance and reliability requirements as cloud computing is most used for IT applications. Different performance requirement leads to different deployment schemes. For example, SFC-RA may require more redundancy to improve reliability. As VMs in cloud computing are hosted in data centers (DCs), the infrastructure of VM placement is homogeneous. However, the infrastructure in SFC-RA is heterogeneous, which can involve optical network, Ethernet, wireless, etc. Therefore, the physical topology in the two problems may be different. Another difference is that SFC has order requirement, which means that the traffic must be steered to traverse through predefined ordered network functions \cite{QuinnNadeau2015}. The main differences of the two problems are summarized in \tablename{ \ref{Comparison-of-SFC-Resource-Allocation and VM Placement}}.
\begin{table}[t]
	\renewcommand{\arraystretch}{1.3}
	\caption{Comparison of SFC-RA and VM placement}
	\label{Comparison-of-SFC-Resource-Allocation and VM Placement}
	\centering
	\footnotesize
	\begin{tabular}{l l l}
		\hline
		Issue&SFC Resource Allocation&VM Placement\\
		\hline
		Objects&VNFs&VMs\\
		Performance&Carrier-grade (five nine)&IT applications\\
		Infrastructure&Heterogeneous&Homogeneous\\
		Order&Required&Not required\\
		\hline
	\end{tabular}
\end{table}

\subsection{Virtual network embedding}
Network Virtualisation decouples infrastructure from services in traditional ISPs, which induces two main entities: Infrastructure Provider (InP) and Service Provider (SP). In this business model, InPs maintain the physical networks (infrastructures), and SPs rent infrastructure from InPs to compose its Virtual Networks (VNs). This approach allows multiple VNs be instantiated in one single physical network, and the VNs are typically independent with each other \cite{ChowdhuryBoutaba2009,SchaffrathWerlePapadimitriouEtAl2009}.
\subsubsection{Introduction to Virtual Network Embedding}
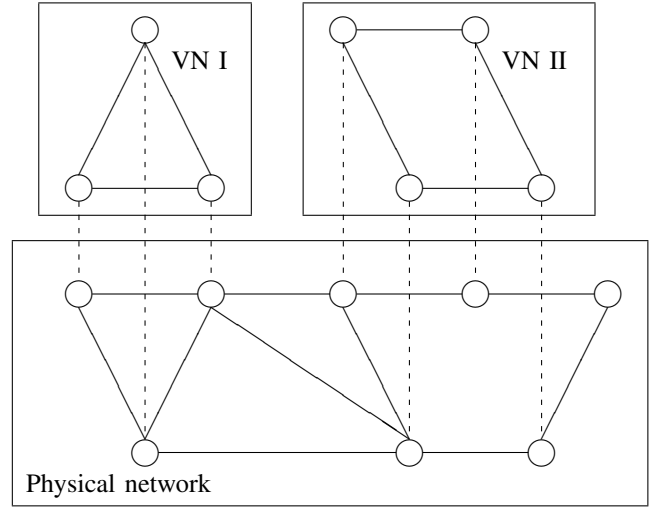
\begin{figure}[t]
	\begin{picture}(\columnwidth,200)(0,0)
	\put(0,0){\framebox(240,100)}
	\put(5,5){Physical network}
	\put(25,80){\circle{10}}
	\put(75,80){\circle{10}}
	\put(125,80){\circle{10}}
	\put(175,80){\circle{10}}
	\put(225,80){\circle{10}}
	\put(50,20){\circle{10}}
	\put(150,20){\circle{10}}
	\put(200,20){\circle{10}}
	\put(30,80){\line(1,0){40}}
	\put(80,80){\line(1,0){40}}
	\put(130,80){\line(1,0){40}}
	\put(180,80){\line(1,0){40}}
	\put(155,20){\line(1,0){40}}
	\put(55,20){\line(1,0){90}}
	\put(25,75){\line(1,-2){25}}
	\put(75,75){\line(-1,-2){25}}
	\put(75,75){\line(3,-2){75}}
	\put(125,75){\line(1,-2){25}}
	\put(225,75){\line(-1,-2){25}}
	\put(25,120){\circle{10}}
	\put(75,120){\circle{10}}
	\put(50,180){\circle{10}}
	\put(25,125){\line(1,2){25}}
	\put(50,175){\line(1,-2){25}}
	\put(30,120){\line(1,0){40}}
	\put(10,110){\framebox(80,80)}
	\put(60,165){VN \uppercase\expandafter{\romannumeral1}}
	\put(125,180){\circle{10}}
	\put(150,120){\circle{10}}
	\put(175,180){\circle{10}}
	\put(200,120){\circle{10}}
	\put(130,180){\line(1,0){40}}
	\put(155,120){\line(1,0){40}}
	\put(125,175){\line(1,-2){25}}
	\put(175,175){\line(1,-2){25}}
	\put(110,110){\framebox(110,80)}
	\put(185,165){VN \uppercase\expandafter{\romannumeral2}}
	\multiput(25,115)(0,-5){6}{\line(0,-1){2}}
	\multiput(75,115)(0,-5){6}{\line(0,-1){2}}
	\multiput(50,175)(0,-5){30}{\line(0,-1){2}}
	\multiput(125,175)(0,-5){18}{\line(0,-1){2}}
	\multiput(175,175)(0,-5){18}{\line(0,-1){2}}
	\multiput(150,115)(0,-5){18}{\line(0,-1){2}}
	\multiput(200,115)(0,-5){18}{\line(0,-1){2}}
	\end{picture}
	\caption{An example of virtual network}
	\label{A-example-of-virtual-network}
\end{figure}

In network virtualisation, a \emph{physical network}, which is also called \emph{substrate network (SN)}, is owned and maintained by an InP. In addition, a physical network consists of physical nodes and physical links. In contrast, a \emph{virtual network} consists of virtual nodes and virtual links. Co-exist of multiple VNs leads to the problem of how the virtual networks be realized by the physical networks, which is the main challenge in network virtualisaton and is called Virtual Network Embedding (VNE) problem. As we mentioned, both VN and SN abstraction are graphs that consist of nodes and links. Therefore, to some extent, VNE problem is finding a subgraph in the physical network topology that is isomorphic with the virtual network topology. \figurename{ \ref{A-example-of-virtual-network}} presents an example of virtual network. However, as a virtual link may be a physical path that consists of several physical nodes and physical links, aforementioned claim is not strictly right. Due to the critical role of VNE in network virtualisation, there are abundant proposals in academic community, e.g., \cite{ChengSuZhangEtAl2011,ZhongbaoSenJunchiEtAl2015,ChowdhuryRahmanBoutaba2009,DietrichRizkPapadimitriou2015}. For example, \citet{ChengSuZhangEtAl2011} apply \emph{Markov Random Walk} model to rank a node based on its resource and toplogical attributes. Then, two algorithms are proposed. First, mapping virtual nodes based on their ranks, and mapping virtual links based on shortest path with unsplittable paths and multi-commodity flow problem with splittable paths. Second, a backtracking VNE algorithm based on breadth-first search. In addition, there are also several surveys about the VNE problem \cite{FischerBoteroBeckEtAl2013,BelbekkoucheHasanKarmouch2012}.

\subsubsection{Relationship between SFC-RA and VNE}
ETSI NFV ISG uses the term Virtual Network Function Forwarding Graph (VNF-FG) \cite{ISG2013-10} instead of Service Function Chaining (SFC) \cite{QuinnNadeau2015} used by IETF SFC WG. In this terminology, VNF-FG consists of a set of virtual network functions and a set of virtual links, which is kind of like a virtual network. Indeed, VNF-FG and Virtual Network are analogous, however, there are also several significant differences.\\
\indent
As Section \ref{Introduction} mentioned, in SFC-RA problem, traffics must flow through predefined ordered network functions. However, VNE do not has such order requirement. In addition, virtual network functions can be shared by multiple flows. However, different virtual networks are typically independent, i.e., a flow of a virtual network \emph{A} do not traverse through the virtual nodes of another virtual network \emph{B} (the flow may traverse the physical nodes that host the virtual nodes of \emph{B}). Especially, the number of different network functions is the outcome of the SFC-RA procedure, however, we know the number of virtual nodes and virtual links of a virtual network in advance. The main differences of the two problems are summarized in \tablename{ \ref{Comparison of SFC Resource Allocation and VNE}}.

\begin{table}[t]
	\renewcommand{\arraystretch}{1.3}
	\caption{Comparison of SFC Resource Allocation and VNE}
	\label{Comparison of SFC Resource Allocation and VNE}
	\centering
	\footnotesize
	\begin{tabular*}{\columnwidth}{l l l}
		\hline
		Issue&SFC Resource Allocation&VNE\\
		\hline
		Objects&VNFs&Virtual nodes and links\\
		Order&Required&Not required\\
		Independence&NFs are not independent&VNs is independent\\
		Number of nodes&Not Know&Know\\
		\hline
	\end{tabular*}
\end{table}

\section{Variants of SFC-RA problem}
\label{Variants-of-SFC-RA-problem}
We have presented a basic mathematical formulation in Section \ref{Basic-mathmatical-model-and-problem-defination}. However, in different variants of SFC-RA problem, modification of the basic formulation is necessary. For example, in delay-sensitive networks, we must consider processing delay caused by VNFs and propagation delay caused by links. Therefore, in such scenario, we should add delay constraints to the model. In addition, in any other variants, we are supposed to modify the basic formulation, too. In this section, we present several variants of SFC-RA problem. We summarize literature about different variants in \tablename{ \ref{Summary-of-SFC-Resource-Allocation}}.
\subsection{Basic}
In basic formulation, we do not consider different variants of SFC-RA problem, e.g., dynamic, online, multiple providers, etc. The basic formulation captures the shared characters that variants of SFC-RA problem have. In this subsection, we summary the basic SFC-RA problem.\\
\indent
For example, \citet{RieraHesselbachZotkiewiczEtAl2015} propose an analytic model for the VNF Forwarding Graph aiming to optimize the execution time of the network services deployed. This work presents a formulation for an optimal SFC-RA; in addition, common economic metrics, performance metrics, etc. are introduced to evaluate and compare different approaches. Therefore, this analytic model can be used by different variants. \citet{GhaznaviShahriarAhmedEtAl2016} model the SFC-RA problem as Mixed Integer Programming (MIP), different from the basic model, the authors consider distributed VNFs and workload balancing, so the formulation is much more complex than basic formulation. Therefore, a local search heuristic is proposed. Similarly, \citet{LuizelliBaysBuriolEtAl2015} decompose the problem into tree phase: (i) placement of VNFs, (ii) assigning VNFs to service requests, (iii) chaining the VNFs, and this insight is similar with the idea presented in \cite{BariChowdhuryAhmedEtAl2015}. The authors use ILP to model this problem and present a heuristic. \citet{CohenLewin-EytanNaorEtAl2015} claim that SFC-RA problem can be reduced to two NP-hard problems, the \emph{facility location problem} and the \emph{generalized assignment problem (GAP)}, which implies that the SFC-RA problem is also NP-hard. Therefore, the authors propose an approximation algorithm based on solving GAP and then rounding the fractional solution computed into an integral solution.

\subsection{Dynamic}
In previous section, we have reviewed a large amount of literature about SFC-RA problem. However, the literature assume that the network is static. In contrast, how to allocate resource at run time is a much complex problem. Although this problem is similar with basic SFC-RA problem, real-time SFC-RA has significant new challenges due to its dynamic features. First, resource that a VNF has may scale due to dynamic traffic. For example, a DPI need less computing resource when the traffic decreases. Second, the QoS demand of VNF may change due to changes of service requests. For example, when an established service request asks for low latency, reallocation of VNFs is required. Third, we should monitor the VNFs for reliability problem. For example, when VNF failures happen, we need reassign VNFs for corresponding service requests\cite{ShiZhangChuEtAl2015}. Therefore, in order to overcome those challenges, we should rethink the SFC-RA problem and propose new solutions.\\
\indent 
\citet{CallegatiCerroniContoliEtAl2015} use OpenFlow to properly steer traffic flows. According to the case study and proof of concept, the authors claim that both layer 2 and layer 3 approaches are functionally viable to implement dynamic SFC. \citet{ShiZhangChuEtAl2015} have the insight that VNFs resources are not allocated simultaneously. Therefore, a preemptive resource allocation strategy is proposed. To realize the strategy, the authors model the SFC-RA problem as \emph{Markov Decision Process (MDP)}. In addition, \emph{Bayesian learning} is used to predict future resource reliability. Leveraging the concept of asynchronous partition \cite{JiaBuyyaChenKhong2005}, the authors propose an algorithm based on MDP.

\subsection{Online}
We sometimes consider service requests that are online, which means that the service requests arrive one by one and are embedded when its arrival. Those solution typically belongs to \emph{Online Algorithms} \cite{Fiat1998}. In such scenario, migration of VNFs may be necessary due to new requests arrival.\\
\indent
\citet{MohammadkhanGhapaniLiuEtAl2015} give a MILP formulation to determine the placement of SFC while minimizing the resource utilization of nodes and links, in order to decrease delay. The highlight of this paper is that the authors develop a heuristic to solve the problem \emph{incrementally}, which support large size instances and can solve problem for incoming flows without impacting existing flows. \citet{LukovszkiSchmid2014} propose a deterministic online algorithm which achieves a competitive ratio of $O(log l)$, where the node capacities are at least logarithmic. In addition, the authors prove that the proposed algorithm is asymptotically optimal in the class of both deterministic and randomized situation. At last, an ILP formulation is presented to show that the problem is NP-complete. \citet{MijumbiSerratGorrichoEtAl2015a} consider the problem of online mapping and scheduling of VNFs. In this situation, each service is created and embedded as its need arises, and VMs can be shared by multiple VNFs. In addition, the authors propose three greedy algorithms and a tabu search-based heuristic.
\subsection{Multiple providers}
\label{Multiple-provicers}
Service functions may be location dependent, for example, proxies and caches should be placed close to the enterprise network. Therefore, a single Network Function Provider (NFP) may not satisfy the location constraints in a service chain, which calls for the coordination of multiple NFPs \cite{DietrichAbujodaPapadimitriou2015}. In addition, the coordination has more benefits, such as improving client experience, cost saving, etc.\\
\indent
Due to NFPs' restrictions in information disclosure, interoperability, etc., new challenges appear in multiple NFPs situation. \citet{AbujodaPapadimitriou2015a} present an architecture, which is called MIDAS, for the coordination of processing setup using a centralized middblebox controller in each NFP. And MIDAS has three basic steps: middlebox signaling, controller chaining, and Multi-Party Computation (MPC). First, MIDAS use a signaling protocol to discover consolidated middleboxes (CoMBs). Then, MIDAS establishes a chain between the controllers of the discovered CoMBs. Then, MIDAS selects CoMBs within each NFP and assigns NFPs via the collaboration of their controllers. At last, upon the CoMB selection, the controller instructs the assigned CoMB(s) to install and configure the required processing modules (PMs). \citet{DietrichAbujodaPapadimitriou2015} first introduce a new service model to simplify the specification of service requests and the estimation of bandwidth demands. In addition, the authors present a topology abstractions tailored to SFC-RA problem, where confidential information of NFPs is concealed. Then a system that embeds the service requests, which is called Nestor, is proposed. Nestor has three main steps. \emph{Graph Rendering}: constructing an abstract topology that spans all NGPs, using the topology abstraction generated by each NFP. \emph{Request Partitioning}: partitioning service requests among NFPs. \emph{NF-subgraph Mapping}: mapping NF-subgraph to the corresponding NFP.

\subsection{Schedule}
In NFV terminology, resource saving is achieved by on-demand resource allocation. In addition, in order to augment resource utilization, it is feasible to use scheduling techniques to allow VNFs to share the resource.\\
\indent
\citet{McGrathRiccobenePetraliaEtAl2015} present a demo that uses resource aware scheduling methods to ensure optimal use of resources and performance in NFV context. \citet{FerrerRieraEscalonaBatalleEtAl2014,FerrerRieraHesselbachEscalonaEtAl2014} formulate the problem of VNF scheduling problem as Resource Constrained Project Scheduling Problem (RCPSP). Similarly, \citet{MijumbiSerratGorrichoEtAl2015a} use MILP to formulate the online virtual function mapping and scheduling problem. And the authors propose three greedy algorithms and a tabu search-based heuristic. \citet{LiQian2015} propose a novel multi-resource fair scheduling algorithm called \emph{Myopia}, which supports multi-resource environments such as NFV. Utilizing the fact that Internet traffic consists of elephant flows and mice flows, Myopia schedules elephant flows precisely and treats mice flows using \emph{First In First Out (FIFO)}. Therefore, Myopia is supposed be a low-complexity and space-efficient packet scheduling algorithm.

\subsection{Mobile network}
\begin{figure}[t]
	\centering
	\includegraphics[width=3in]{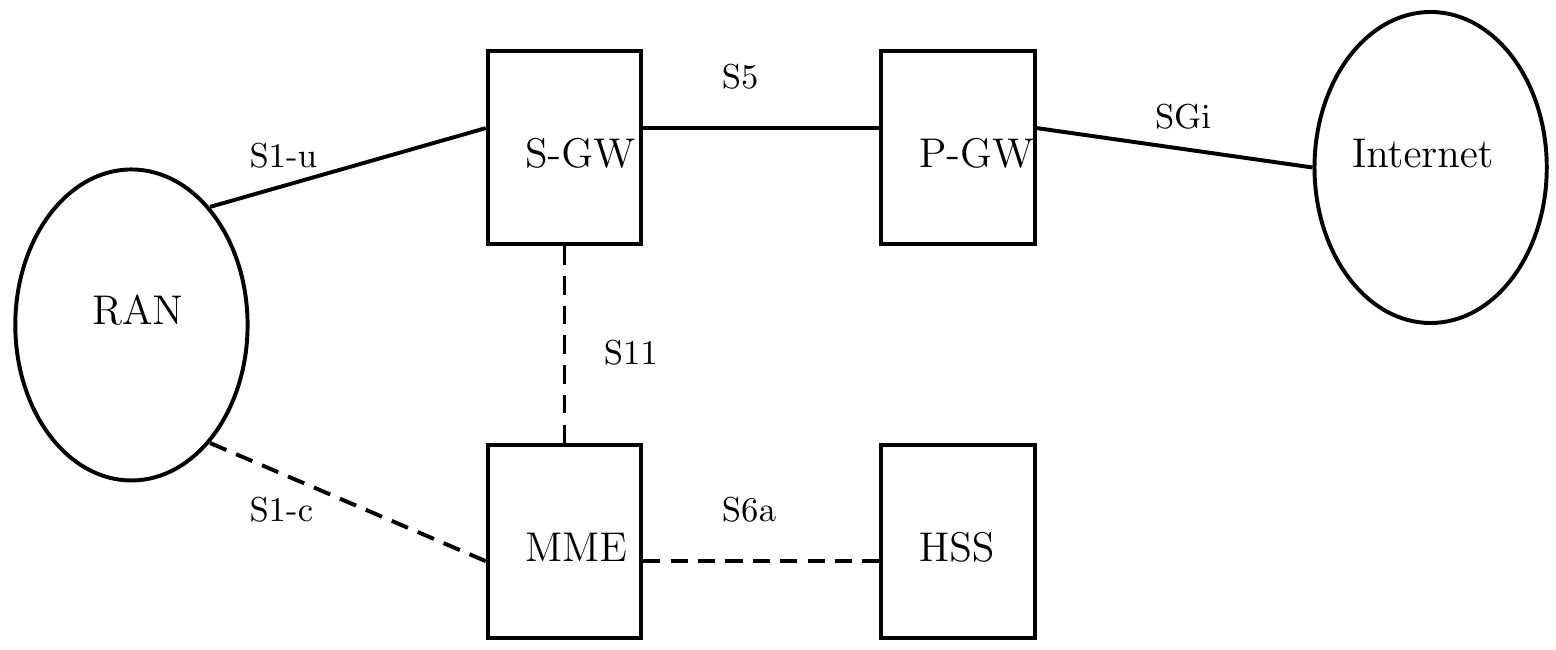}
	\caption{Evolved packet core}
	\label{evolved_packet_core}
\end{figure}

Mobile network have a lot of differences from the Internet, which should be considered in service chain deployment. For example, comparing with fixed network, nodes in wireless access network have one more kind of resource, i.e. radio resource \cite{RiggioBradaiHarutyunyanEtAl2016}. Nevertheless, in Evolved Packet Core (EPC), nodes do not have radio resource. But we typically assume that every nodes have all kinds of resources in fixed network. In EPC, there are fixed kinds of network functions and service function chains, which is more simple than fixed network. An example of EPC is presented in \figurename{ \ref{evolved_packet_core}}. In addition, user mobility is one of the biggest differences between fixed and mobile network. The mobility of user data may cause relocation of Serving Gateway (SGW) or Mobility Management Entity (MME), which incurs cost and impact the overall QoE \cite{TalebKsentini2013}. In addition, mobile network has more service functions than Internet, e.g., \emph{TCP Optimizer}, \emph{Video Optimizer}, \emph{Header Enrichment}, etc., which should be thought over during resource allocation.\\
\indent
\citet{TalebCoriciParadaEtAl2015} demonstrate the feasibility of on-demand creation of cloud-based elastic mobile core networks, and present the requirements and challenges of EPC as a Service (EPCaaS). Then the authors discuss several implementation options. \citet{BaumgartnerReddyBauschert2015} propose a MILP to model the virtual mobile core network embedding with respect to latency bounds. \citet{TalebKsentini2013} consider the need for avoiding or minimizing the relocation of gateway due to user mobility and propose an efficient network function placement algorithm for the realization of mobile cloud. \citet{TalebBagaaKsentini2015} consider two conflicting objectives of VNF placement in mobile core network, i.e. the guarantee of QoE via closer placement of data anchor gateway to user and the avoidance of the relocation of mobility anchor gateway via the placement of VNFs far enough from users. And three solutions are presented, two solutions prefer one objective to another, while the third one try to find a fair trade-off between the two objectives via Nash theory.

\begin{table*}[t!]
	\renewcommand{\arraystretch}{1.3}
	\caption{Summary of Service Function Chaining Resource Allocation}
	\label{Summary-of-SFC-Resource-Allocation}
	\centering
	\rowcolors*{2}{white}{black!10}
	\resizebox{.89\textwidth}{!}{
	\begin{threeparttable}			
		\begin{tabular}{>{\columncolor[gray]{.5}[-30pt]}l l l l l}
			\hline
			Category&Reference&Model&Optimization&Contribution\\		
			\hline \noalign{\smallskip}
			Basic&\citet{RieraHesselbachZotkiewiczEtAl2015}&MILP&N/A\tnote{1}&Introduces an analytic model and several metrics to evaluate the approaches\\
			&\citet{GhaznaviShahriarAhmedEtAl2016}&MIP&Heuristic &Distributed VNF deployment and workload balancing\\
			&\citet{LuizelliBaysBuriolEtAl2015}&ILP&Heuristic&Analysis the topological components of SFC requests\\
			&\citet{NemethCzentyeVaszkunEtAl2015}&N/A&Heuristic&Proposes a fine-tuning of real-time parameterizable orchestration algorithm\\
			&\citet{GuerzoniDespotovicTrivisonnoEtAl2014}&MIP&N/A&Presents a comprehensive analysis and model of reliability requirements\\
			&\citet{BariChowdhuryAhmedEtAl2015}&ILP&Heuristic&Introduces network transformation to simplify the formulation\\
			&\citet{EramoMiucci}&ILP&Heuristic&Proposes a consolidation algorithm for consumption minimization\\
			&\citet{GuptaHabibChowdhuryEtAl2015}&ILP&N/A&Presents a comprehensive mathematical model\\
			&\citet{CohenLewin-EytanNaorEtAl2015}&ILP&Approximation&Provides approximation algorithm with theoretically proven performance\\
			&\citet{XiaShirazipourZhangEtAl2014}&BIP&Heuristic&Considers conversion cost in packet/optical datacenters\\
			&\citet{MehraghdamKellerKarl2014}&MIQCP&N/A&Presents a context-free language to formalize the service requests\\
			&\citet{AddisBelabedBouetEtAl2015}&MILP&Heuristic&Considers flow compression/decompression constraints\\
			&\citet{MoensDeTurck2014}&MILP&N/A&Considers the scenario of coexisting with physical functions\\
			&\citet{HubermanSharma2016}&CAP\tnote{2}&Exact&Leverages \emph{comparative advantage} to place VNFs\\
			&\citet{LinZhouTornatoreEtAl2016}&MIP, GT\tnote{3}&Heuristic&Presents an algorithm based on \emph{iterative weakly dominated elimination}\\
			&\citet{BruschiCarregaDavoli2016}&GT&N/A&An energy-aware Game-Theory-based solution for VNFs placement\\
			&\citet{MartiniPaganelliCappaneraEtAl2015}&RCSP\tnote{4}&N/A&A layered structure that ensures the order of VNFs specified in the request\\
			&\citet{SuksomboonFukushimaHayashiEtAl2015}&IP\tnote{5}&Heuristic&A framework to identify which VNF is worth to be outsourced to the cloud\\
			&\citet{YoshidaShenKawabataEtAl2014}&N/A&Metaheuristic&Considers multi-objective VNF placement\\
			&\citet{RostSchmid2016}&IP&Approximation&A polynomial algorithm for service chain embedding\\
			&\citet{GemberGrandlAnandEtAl}&N/A&Heuristic&Presents an orchestration layer for VNFs in clouds\\
			&\citet{SahhafTavernierColleEtAl2015}&N/A&Heuristic&Considers VNF placement supporting service decomposition\\
			&\citet{SahhafTavernierRostEtAl2015}&ILP&Heuristic&Improvement of \cite{SahhafTavernierColleEtAl2015}\\
			&\citet{FanGuanRenEtAl}&N/A&Approximation&Availability guaranteed SFC mapping using geographic redundancy\\
			\noalign{\smallskip} \hline \noalign{\smallskip}	
						
			Dynamic&\citet{ShiZhangChuEtAl2015}&MDP&Heuristic&Adopts \emph{Markov Decision Process} to model the problem\\
			&\citet{CallegatiCerroniContoliEtAl2015}&N/A&N/A&A OpenFlow Based implement of dynamic chaining of VNFs\\
			&\citet{ClaymanMainiGalisEtAl2014}&N/A&N/A&Presents an architecture that ensures the automatic placement of VNFs\\
			&\citet{GhaznaviKhanShahriarEtAl2015}&ILP&Heuristic&Elastic VNF placement\\
			\noalign{\smallskip}
			\hline 
			\noalign{\smallskip}
			
			Online&\citet{MohammadkhanGhapaniLiuEtAl2015}&MILP&Heuristic&Develops a heuristic to solve the problem incrementally\\
			&\citet{LukovszkiSchmid2014}&ILP&Deterministic&Proposes a deterministic online algorithm\\	
			&\citet{MijumbiSerratGorrichoEtAl2015a}&N/A&(Meta-)Heuristic&Firstly formulates the problem of online mapping and scheduling of VNFs\\
			&\citet{WangWuLeEtAl2016}&IP&RA\tnote{6}, Heuristic&An online algorithm that dynamically place VNFs in DCs\\
			&\citet{MaMedinaPan2015}&IP&Exact&Presents a polynomial algorithm that placement VNFs one by one\\
			\noalign{\smallskip}
			\hline
			\noalign{\smallskip}
			
			Multiple&\citet{AbujodaPapadimitriou2015a}&N/A&N/A&An architecture for SFC deployment across multi-providers\\
			Providers&\citet{DietrichAbujodaPapadimitriou2015}&ILP&Heuristic&An allocation approach among multi-providers that respects providers' privacy\\
			&\citet{AbujodaPapadimitriou2016}&N/A&N/A&Ensures competitive pricing among multiple providers\\
			&\citet{BhamareJainSamakaEtAl2015}&ILP&Heuristic&VNF placement algorithm that minimize total delay in multi-cloud scenario\\
			\noalign{\smallskip}
			\hline
			\noalign{\smallskip}
			
			Schedule&\citet{FerrerRieraEscalonaBatalleEtAl2014}&ILP&N/A&Provides the first formalisation of the VNF scheduling problem\\
			&\citet{MijumbiSerratGorrichoEtAl2015a}&N/A&Heuristic&Firstly formulates the problem of online mapping and scheduling of VNFs\\
			&\citet{McGrathRiccobenePetraliaEtAl2015}&N/A&N/A&A demo using resource aware scheduling methods in NFV\\
			&\citet{LiQian2015}&N/A&Schedule&A novel low-complexity and space-efficient packet scheduling algorithm\\
			\noalign{\smallskip}
			\hline
			\noalign{\smallskip}
			
			Mobile&\citet{BaumgartnerReddyBauschert2015}&MILP&N/A&Improvement of \cite{BaumgartnerReddyBauschert2015a} that considers latency bounds\\
			Network&\citet{BaumgartnerReddyBauschert2015a}&ILP&N/A&A model for VNF placement and topology optimization in mobile core network\\
			&\citet{YousafLoureiroZdarskyEtAl2015}&N/A&Heuristic&Cost analysis of two heuristic approaches of vEPC deployment\\
			&\citet{TalebKsentini2013}&ILP&Heuristic&Approach for gateway relocation avoidance-aware VNF placement\\
			&\citet{RiggioBradaiHarutyunyanEtAl2016}&ILP&Heuristic&Improvement of \cite{RiggioBradaiRasheedETAL2015}\\
			&\citet{TalebBagaaKsentini2015}&ILP&Exact&An approach for VNF placement in mobile network considering user mobility\\
			&\citet{BagaaTalebKsentini2014}&NO\tnote{7}&Heuristic&An approach that focuses on the data anchor (\emph{PDN-GW}) virtualization\\
			&\citet{RiggioBradaiRasheedETAL2015}&ILP&Heuristic&Considers VNF placement in radio access networks\\
			&\citet{BastaKellererHoffmannEtAl2014a}&ILP&N/A&Formulation of virtual GW placement in SDN and NFV environment \\
			&\citet{MijumbiSerratGorrichoEtAl2015b}&BILP&Heuristic&Considers BBU placement and assignment in virtual radio access networks\\
			\noalign{\smallskip}
			\hline
			\noalign{\smallskip}
			Data&\citet{FangxinRuilinZhuEtAl2015}&IP&Heuristic&Bandwidth guaranteed VNF placement and scaling in DC\\
			Center\tnote{8}&\citet{HerkerAnKiessEtAl2015}&N/A&Heuristic&Availability of VNF placement in data center\\
			&\citet{MedhatCarellaCEtAl2015}&N/A&Heuristic&A VNF placement algorithm provides tradeoff between delay and load balancing\\
			\noalign{\smallskip}
			\hline
		\end{tabular}
		\begin{tablenotes}
			\footnotesize

			\item[1]N/A is the abbreviation of \emph{Not Applicable}
			\item[2]CAP is the abbreviation of \emph{Comparative Advantage Problem}
			\item[3]GT is the abbreviation of \emph{Game Theory}
			\item[4]RCSP is the abbreviation of \emph{Resource Constrained Shortest Path}			
			\item[5]IP is the abbreviation of \emph{Integer Programming}	
			\item[6]RA is the abbreviation of \emph{Randomized Algorithm}
			\item[7]NO is the abbreviation of \emph{Nonlinear Optimization}
			\item[8]We discuss variant about data center in Section \ref{SFC-RA-in-data-center}
		\end{tablenotes}
	\end{threeparttable}
	}
\end{table*}

\section{Future research directions}
\label{Future-research-directions}
After our survey of literature about SFC-RA problem, we have a full view about this problem. Although there are a huge quantity of papers have been published, which are discussed in previous sections, SFC-RA is still in early stage. The works we mentioned before need more comprehensive study. Moreover, there still remain significant research directions that should be investigated. This section discusses future research directions.

\subsection{Resiliency}
Due to the development of Cloud Computing and Network Virtualization, virtualized data centers have been deployed in cloud providers' infrastructure. However, in the telecommunication domain, there are no widespread deployments yet. One of the most significant differences of IT and telecommunication is performance requirement: telecommunication have a performance requirement of \emph{five nines}, while the IT domain does not have such rigid requirement. Other performance requirements of telecommunication are automatic service recovery, limited amount users that are bothered by outages, etc. \cite{ISG2015-01}. While deal with resilience problem in SFC, solutions can make use of MAC- or IP-level redundancy mechanisms such as Virtual Router Redundancy Protocol (VRRP). Also, particularly for SF failures, load balancers co-located with the SFF or as part of the service function delivery mechanism can provide such robustness \cite{HalpernPignataro2015}.\\
\indent
\citet{LeeShin2015} consider the Service Function Path recovery due to service functions failures. The authors propose a scheme that temporally recovers the traffic path from failures by shifting the responsibility of a service function in failure to another service function with data-plane signaling. In addition, the authors intend to study a algorithm of selecting remote SFFs in optimal. \citet{FanGuanRenEtAl} consider the problem of availability-aware SFC mapping. The authors propose a novel enhanced \emph{Joint Protection (JP)} approach which is better than traditional \emph{Dedicated Protection (DP)} and \emph{Shared Protection (SP)}. After proving the NP-hardness of the availability-aware SFC mapping, the authors propose an approximation algorithm with a theoretical lower bound. \citet{Schx00FllerEtAl2013} introduce an architecture to deploy VNF on cloud infrastructure. The authors also leverage the concept of \emph{availability zone} in OpenStack to ensure service deployment resilience.\\
\indent
Especially, there are a lot of literature about \emph{resilient} Virtual Network Embedding, e.g., \cite{RahmanBoutaba2013,JarraySongKarmouch2013,AilingYingLuomingEtAl2013}, from which are worth drawing inspiration. For example, \citet{RahmanBoutaba2013} consider that the InP network does not remain operational at all times. In order to solve VNE problem in such scenario, the authors propose a proactive and a hybrid policy heuristic. And the hybrid policy is based on a fast re-routing strategy and utilizes a pre-reserved quota for backup on each physical link.

\subsection{Distributed SFC-RA}
Most literature we mentioned presents centralized approaches which means that a single node computes all the resource allocation. Centralized approaches suffer from the problem of scalability, further more, centralized algorithms may be infeasible in some situations, such as multiple providers and multiple administrative domains. Therefore, distributed approach seem to be a feasible solution to overcome those challenges. In addition, ETSI NFV ISG has a PoC proposal about distributed NFV, which is named \emph{Multi-vendor Distributed NFV}\footnote[1]{Online avaliable: \url{http://nfvwiki.etsi.org/index.php?title=Multi-vendor_Distributed_NFV}}.\\
\indent
As we discussed in Section \ref{Multiple-provicers}, we are supposed to consider SFC-RA problem in the situation of multiple providers, in which multiple network function providers cooperate to compose a service chain, due to several reasons such as location constraints and cost saving. In such scenario, distributed SFC-RA is necessary as we can not compute all the VNF embedding in one single node. Therefore, Nestor \cite{DietrichAbujodaPapadimitriou2015}, MIDAS \cite{AbujodaPapadimitriou2015a}, and DistNSE \cite{AbujodaPapadimitriou2016} present distributed approaches for SFC-RA. In aforementioned situation, different network function providers are homogeneous, which means the providers have same network functions. \citet{RosaSantosRothenberg2015} consider a heterogeneous environment, in which service chains span several administrative domains, i.e. data centers, carriers, and CPEs. The authors claim that the optimization of VNF placement within different locations have several advantages, for example, maximizing the QoE by bring VNFs closer to users, and minimizing the costs by consolidating more VNFs in data centers. In addition, the authors discuss three use cases of multi-domain distributed NFV, i.e. management and orchestration (MANO), bandwidth negotiation, and reliability. In conclusion, we need novel algorithms to deal with SFC-RA problem in distributed NFV environment for optimization considers.

\subsection{SFC-RA in data center}
\label{SFC-RA-in-data-center}
There are two primary types of traffic in data center (DC) context: \emph{north-south} and \emph{east-west}. North-south traffic originates from outside the DC and is typically associated with users. East-wast traffic originates from one DC, and end with another DC, which is the predominant traffic in data center today. Therefore, there are two kinds of SFC in DC context: intra-DC SFCs and inter-DC SFCs \cite{KumarTufailMajeeEtAl2016}. One of the biggest features of DC is the regularity of physical topology, whose typical architectures are \emph{multi-tier tree}, \emph{fat-tree}, \emph{BCube}, \emph{DCell}, etc. \cite{HerkerAnKiessEtAl2015}. In addition, DCs are typically homogeneous, which means that servers have the same capacity of computing, storage, and communication. Leveraging the features of DC, we may design SFC-RA algorithms that are suitable for DC context. \\
\indent
For example, \citet{HerkerAnKiessEtAl2015} consider network functions chain embedding in DCs. Leveraging different backup strategies and algorithms, resilient service chain embedding in DCs is presented. Furthermore, the authors investigate the DC architecture impacts on availability of SFC embedding. From the results of the paper, \emph{2-tier tree} topology is the best topology for achieving high availability for SFC embedding. \citet{FangxinRuilinZhuEtAl2015} consider bandwidth guaranteed VNF placement and scaling in DC. Leveraging the tree-like topology of DC networks, the authors propose an on-line heuristic algorithm that achieves approximation optimal allocation.

\section{Conclusion}
\label{Conclusion}
Service Function Chaining Resource Allocation is a crucial problem to be solved for deploying service function chains in NFV environment. The problem of computing optimal allocation solutions is NP-hard, hence unsolvable for large problem size. Therefore, there is strong demand for efficient algorithms to solve the problem. A huge quantity of approaches have been proposed in the literature, so far.\\
\indent
In this paper, we present a survey of current work in this research direction. A formal formulation and several optimization strategies were presented. We discussed the relationships of SFC-RA problem with Virtual Machine placement problem and Virtual Network Embedding problem. Then we presented several variants, and summarized different approaches that solve the problem.\\
\indent
There are great opportunities for future work in this area. High resiliency requirement of Service Function Chaining is one of the main features, which should be paid high attention during the deployment phase. It is worth noting that there is few literature concern distributed SFC-RA problem. This is a promising point for future work, since centralized approaches suffer from scalability problem and may be infeasible in some situation, e.g., multiple providers. Moreover, leveraging the exclusive features of data center, it is possible to design more suitable algorithms for data center environment.

\bibliographystyle{IEEEtranN}
{\footnotesize\bibliography{Service_Function_Chaining_Resource_Mapping-A_Survey}}

\end{document}